\documentstyle[prl,eqsecnum,aps]{revtex}
\twocolumn
\begin{document}
\author{Jian-Qi Shen \\{\it E-mail address}: jqshen@coer.zju.edu.cn}
\address{Zhejiang Institute of Modern Physics and Department of Physics,
Zhejiang University, Spring Jade, \\Hangzhou 310027, P. R. China}
\date{\today }
\title{On some weak-gravitational effects}
\maketitle

\begin{abstract}
 ${\mathcal ABSTRACT}$

Some physically interesting weak-gravitational effects and
phenomena are reviewed and briefly discussed: particle geometric
phases due to the time-dependent spin-rotation couplings,
non-inertial gravitational wave in rotating reference of frame,
hyperbolical geometric quantum phases and topological dual mass as
well.
\end{abstract}
\section{$\mathcal Introduction$}
On the basis of the theoretical and experimental works concerning
the spin-rotation coupling by Mashhoon {\it et al.}\cite{27,32},
we investigate some relativistic quantum gravitational effects
associated with gravitomagnetic fields: the {\it geometric quantum
phase factor} of a spinning particle interacting with
time-dependent gravitomagnetic fields and the {\it non-inertial
gravitational wave} in time-dependent rotating reference frame. By
means of neutron-gravity interferometry experiment, this geometric
phase factor due to time-dependent spin-rotation coupling can be
applied to obtaining information on small fluctuations of Earth's
rotating frequency\cite{20}. Mashhoon's spin-rotation coupling is
a static case, while the interaction of the intrinsic spin of a
particle with the non-inertial gravitational wave is its mobile
realization. With the improvements in precise-measuring
instruments, particularly the laser-interference technology, it
becomes possible for investigating quantum mechanics in
weak-gravitational fields, which provides tests of Einstein theory
of general relativity in quantum regime.

Both {\it geometric phases}\cite{1,2} of wave function in Quantum
Mechanics and {\it gravitomagnetic charge} (topological dual
charge of mass)\cite{Bell,Nouri,Shen1,Shen2} in the general theory
of relativity reveal Nature's geometric or global properties.
Differing from the dynamical phase, geometric phase depends only
on the geometric nature of the pathway along which the quantum
system evolves\cite{3,4}. Here we show the existence of the
hyperbolical geometric quantum phase that is different from the
ordinary trigonometric geometric quantum phase\cite{5}.
Gravitomagnetic charge (dual mass) is the gravitational analogue
of magnetic monopole in Electrodynamics\cite{6}; but, as we will
show here, it possesses more interesting and significant features,
{\it e.g.}, it may constitute the {\it dual matter} that has
different gravitational properties compared with mass. In order to
describe the space-time curvature due to the {\it topological dual
mass}, we construct a dual Einstein's tensor. Further
investigation shows that gravitomagnetic potentials caused by {\it
dual mass} are respectively analogous to the trigonometric and
hyperbolic geometric phase. The study of the geometric phase and
dual mass provides a valuable insight into the time evolution of
quantum systems and the topological properties in General
Relativity.
\section{$\mathcal Gravitomagnetic$ $\mathcal fields,$ $\mathcal spin-rotation$ $\mathcal couplings$ $\mathcal and$ $\mathcal geometric$ $\mathcal phases$}
According to the equivalence principle in general relativity, the
nature of spin-rotation coupling is merely the interaction between
the gravitational spin-magnetic moment and the gravitomagnetic
field. By considering the Doppler effect of a light signal in the
non-inertial reference frame rotating relative to the fixing
reference frame, Mashhoon obtained the Hamiltonian of the
spin-rotation coupling, which is of the form
$H=\vec{\omega}\cdot{\bf S}$ with $\vec{\omega}$ and ${\bf S}$
respectively denoting the rotating frequency of the rotating frame
and the intrinsic spin of a photon. In the framework of general
relativity, however, we transformed the Kerr metric from the
fixing reference frame to the rotating frame, and therefore
obtained the exterior gravitomagnetic potential (one of the metric
components $g_{0\varphi}$) of the spherically symmetric rotating
body as follows\cite{20}
\begin{equation}
g_{0\varphi}\simeq\frac{2aGM\sin^{2}\theta}{c^{2}r}+\frac{2\omega
r^{2}\sin^{2}\theta}{c},      \label{eq1}
\end{equation}
which is calculated by using the weak-field approximation, where
$c$ and $G$ are the speed of light and the gravitational constant,
respectively; $a$ is so defined that $ac$ represents the angular
momentum per unit mass of this gravitating body, and $(r, \theta,
\varphi)$ stands for the displacements of the spherical coordinate
system fixed in the rotating reference frame. It follows that the
first term on the right handed side in Eq.(\ref{eq1}) is related
to the gravitational constant $G$ and is vanishing when the radial
coordinate $r\rightarrow \infty$, whereas the second term tends to
infinity rather than vanishes since it arises from the coordinates
transformation in terms of the equivalence principle. It is of
interest that the second term in Eq.(\ref{eq1}) also gives rise to
a gravitomagnetic field whose strength is just the rotating
frequency, $\vec{\omega}$, of the rotating frame with respect to
the fixing reference system. It is well known that the magnetic
field results in the Lorentz force acting on a charged particle,
in the similar fashion, the {\it non-inertial gravitomagnetic
field} $\vec{\omega}$ also yields a fictitious Coriolis force
acting on a moving particle. In view of above discussion, one can
also obtain the same Hamiltonian of spin-rotation coupling by
using the Dirac equation with spin connection in curved
spacetimes. Essentially, we can draw a conclusion that the
interaction of the gravitational magnetic moment and the
gravitomagnetic field involves spin-rotation coupling. The
spin-rotation coupling leads to the inertial effects of the
intrinsic spin of a particle, for instance, although the
equivalence principle still holds, the universality of Galileo's
law of freely falling particles is violated, provided that the
spin is polarized vertically up or down in the non-inertial
frame\cite{27}.

Since there exists observational evidence for the coupling of
spin-$\frac{1}{2}$ particle with the rotation of the Earth, we
suggest a geometric effect in the coupling of the neutron spin
with the time-dependent rotation of the Earth. Since the analogy
can be drawn between gravity and electromagnetic force in some
aspects, Aharonov and Carmi proposed the geometric effect of
vector potentials of gravity, and Anandan, Dresden and Sakurai
{\it et al.} proposed the quantum-interferometry effect associated
with gravity\cite{Anandan,Dresden,Overhauser,Werner}. What they
investigated is now termed the Aharonov-Carmi effect, which is the
gravitational analogue to the Aharonov-Bohm effect in
electrodynamics that appears as a geometric quantum phase factor
({\it i.e.}, Berry's phase factor) in the wave function of an
electron moving along a closed path surrounding a magnetic flux.
Although the spin-rotation coupling and Aharonov-Carmi effect have
the same origin, namely, both arise from the presence of the
Coriolis force ({\it i.e.}, the interaction of the motion of the
particle with the non-inertial gravitomagnetic field,
$\vec{\omega}$), we argue that the Aharonov-Carmi effect mentioned
above does not comprise the new geometric effect due to the
time-dependent rotation of the non-inertial frame.

It is well known that the geometric phase factor appears in the
quantum systems whose Hamiltonian explicitly depends on time or
possesses evolution parameters. Differing from the dynamical phase
which is related to the energy, frequency or velocity of a
particle or a quantum system, geometric phase is dependent only on
the geometric nature of the pathway along which the system
evolves, which reflects the global and topological properties of
evolution of the quantum systems. By making use of the
Lewis-Riesenfeld invariant theory and the invariant-related
unitary transformation formulation\cite{7}, we compute the
geometric phase that results from the neutron spin-rotation
coupling in which the rotating frequency is {\it time-dependent}.
The result may be written as follows
\begin{equation}
\phi_{\pm}=\pm\frac{1}{2}\int_{0}^{t}\frac{{\rm
d}\varphi(t')}{{\rm d}t'}\left[1-\cos \theta
\left(t'\right)\right]{\rm d}t'                 \label{eq2}
\end{equation}
with $(\theta, \varphi)$ being the angle displacements over the
parameters space of the conserved invariant, an operator whose
eigenvalue is {\it time-independent}; $\pm$ corresponding to the
neutron spin polarized vertically up and down. For the case of the
adiabatic limit in which $\theta$ is constant, the geometric phase
in one cycle over the parameter space is then
$\phi_{\pm}=\pm\frac{1}{2}\cdot 2\pi(1-\cos \theta)$, where
$2\pi(1-\cos \theta)$ is the expression for the solid angle which
presents the geometric properties of time evolution of this
spin-rotation system.

At present, investigation of geometric phase factor is an
important direction in atomic and molecular physics, quantum
optics, condensed matter physics and molecular reaction chemistry
as well. Geometric phase factor has many applications in various
branches of physics, say, in the case of this spin-rotation
coupling, a potential application can be suggested. Since this
geometric phase reveals the time evolution of $\vec{\omega}(t)$,
information on the Earth's variations of rotation will be obtained
by measuring the geometric phase of the oppositely polarized
neutrons through the neutron interferometry experiment.
\section{$\mathcal Non-inertial$ $\mathcal gravitational$ $\mathcal wave$ $\mathcal in$ $\mathcal rotating$ $\mathcal reference$ $\mathcal of$ $\mathcal frame$}
Note that Mashhoon's spin-rotation or spin-gravity coupling is
only confined within the static case in which the rotating
frequency is independent of time. Here we propose another
interaction where the intrinsic spin of a particle is coupled to a
propagating gravitomagnetic field. Einstein theory of General
Relativity predicts that the accelerated mass produces the
propagation of the space-time curvature which is termed the
gravitational wave. Detecting and investigating these space-time
ripples is one of the leading areas in astrophysics and cosmology.
Because of the weakness of the gravitational radiation,
gravitational wave has not been detected up to now even by means
of both resonant-mass detectors and laser-interferometric
detectors. In this paper, however, we suggest the concept of {\it
non-inertial gravitational wave}, which does not relate to the
smallness of the gravitational constant. Further detailed reasons
are illustrated in the following. It follows from Eq.(\ref{eq1})
that, in the rotating reference frame, the rotation yields
additional space-time curvature (expressed by $\frac{2\omega
r^{2}\sin^{2}\theta}{c}$) which is independent of the
gravitational constant, $G$. If, therefore, the rotating frequency
$\vec{\omega}$ varies with time, the variation of the space-time
curvature (the disturbance of the gravitational field) in the
rotating reference frame propagates outwards in the form of wave
motion. Such a propagating disturbance is referred to as a {\it
non-inertial gravitational wave} (NGW). By ignoring some
negligible and higher-order terms, the wave equation of motion
involving source terms is readily obtained
\begin{equation}
\left(\frac{1}{c^{2}}\frac{\partial^{2}}{\partial
t^{2}}-\nabla^{2}\right)h_{mn}+\frac{16\pi
G}{c^{4}}S_{mn}=\eta_{ik}\frac{\partial^{2}}{\partial
t^{2}}\left(a^{i}_{m}a^{k}_{n}\right),              \label{eq3}
\end{equation}
where $h_{mn}$ and $\eta_{ik}$ denote the amplitudes of
gravitational wave and the metric tensor of the flat Minkowski
spacetime, respectively; $a^{i}_{m}$ and $a^{k}_{n}$ are the
matrix elements of coordinate transformation from the fixing frame
to the rotating frame; $S_{mn}=T_{mn}-\frac{1}{2}\eta_{mn}T$ with
$T_{mn}$ being the energy-momentum tensor of matter and
$T=\eta^{mn}T_{mn}$. By using the low-motion weak-field
approximation, further analysis shows that there exist only two
kinds of amplitudes of NGW, {\it i.e.}, $h_{00}$ and
$h_{0\varphi}$ in the time-dependent rotating reference frame, and
the expressions of the source term on the right handed side of
Eq.(\ref{eq3}) are respectively of the form $-r^{2}\sin^{2}\theta
\frac{\partial^{2}}{\partial
t^{2}}\left(\frac{\omega^{2}}{c^{2}}\right)$ and
$r^{2}\sin^{2}\theta \frac{\partial^{2}}{\partial
t^{2}}\left(\frac{\omega}{c}\right)$. Note, however, that this NGW
depends on the non-inertial reference frame and differs from the
standard gravitational wave predicted by Einstein.

Experimental evidence for spin-rotation coupling exists in the
microwave and optical regimes via the phenomenon of frequency
shift of polarized radiation. It is believed that with the
foreseeable improvements in detecting technology, the interaction
between the intrinsic spin of fundamental particles and NGW can
also be detected. Since the existence of spin-rotation coupling
can be observed only from the non-inertial frame, NGW is thus
defined to be such space-time ripples associated with the
non-inertial reference frame. It is therefore apparent that
investigation of NGW is of much physical interest since its
intensity is no longer dependent on the gravitational wave.

Although it possesses the non-inertial properties, this NGW can be
coupled to matter. It is shown in what follows that the coupling
coefficient depends on the gravitational constant, $G$.

In London's electrodynamics of superconductivity, the velocity
${\bf v}$ of the superconducting electrons and the magnetic
potentials ${\bf A}$ satisfies $\frac{\partial}{\partial t}{\bf
v}=-\frac{e}{m_{\rm e}}\frac{\partial}{\partial t}{\bf A}$ which
leads to the self-induced charge current and thus provides photons
field with an effective mass term in field equation or Lagrangian
of electrodynamics. This procedure is equivalent to the mechanism
of spontaneously symmetrical breaking. In the theory of gravity,
the similar phenomenon exists since the momentum density
$\rho_{\rm m}{\bf v}$ is conserved before and after the particles
scattering, which is in analogy with the case in the
superconductor where the electric current density is also
conserved as the scattering is inhibited for Cooper pairs. Note
that the Einstein's equation of gravitational field under the
low-motion weak-field approximation is formally analogous to the
Maxwell equation of electromagnetic field, then in the case of
gravitating matter, one can arrive at the similar relation
$\frac{\partial}{\partial t}{\bf v}=c\frac{\partial}{\partial
t}{\bf h}$ between the velocity of particles and the
gravitomagnetic vector potentials ${\bf h}$, where ${\bf
h}=(g^{01}, g^{02}, g^{03})$. From the point of view of the
low-motion weak-field approximation, the {\it self-induced mass
current} results from this coupling of gravity to matter and then
enables the gravitational field to possess an {effective rest mass
squared}, $m_{\rm g}^{2}=-\frac{\hbar^{2}}{c^{4}}(8\pi G\rho_{\rm
m})$, where $\hbar$ and $\rho_{\rm m}$ denote the Planck constant
and the mass density of matter, respectively. Apparently, the NGW
is not merely the non-inertial effect, for the effective mass of
NGW is related to the gravitational constant, $G$.
\section{$\mathcal Hyperbolical$ $\mathcal geometric$ $\mathcal quantum$ $\mathcal phases$ $\mathcal and$ $\mathcal topological$ $\mathcal dual$ $\mathcal mass$}
Geometric phase exists in time-dependent quantum systems or
systems whose Hamiltonian possesses evolution
parameters\cite{7,8}. As is well known, the dynamical phase of
wave function in Quantum Mechanics is dependent on dynamical
quantities such as energy, frequency, coupling coefficients and
velocity of a particle or a quantum system, while the geometric
phase is immediately independent of these physical quantities.
When Berry found that the wave function would give rise to a
non-integral phase (Berry's phase) in quantum adiabatic
process\cite{1}, geometric phase attracts attention of many
physicists in various fields such as gravity theory\cite{9,10},
differential geometry\cite{4}, atomic and molecular
physics\cite{11}, nuclear physics\cite{12}, quantum
optics\cite{13,14,15}, condensed matter physics\cite{16,17,18,19}
and molecular-reaction chemistry\cite{11} as well. In many simple
quantum systems such as an electron possessing intrinsic magnetic
moment interacting with a time-dependent magnetic field (or a
neutron spin interacting with the Earth's rotation\cite{20}), a
photon propagating inside the curved optical fiber\cite{5,15}, and
the time-dependent Jaynes-Cummings model describing the
interaction of the two-level atom with a radiation field\cite{21},
geometric phase is often proportional to $2\pi(1-\cos \theta)$,
which equals the solid angle subtended by a curve with respect to
the origin of parameter space\footnote{The solid angle subtended
by the curve trace on the cone, at the center, shows the
topological meanings of geometric phase. Geometric phase is absent
in the quantum system when its Hamiltonian is independent of time.
Geometric phase of many simple physical systems in the adiabatic
process can be expressed in terms of the solid angle over the
parameter space, {\it e.g.}, the wave function of a photon
propagating inside the non-coplanar curved optical fiber obtains
this topological phase, where the parameter space is just the
momentum ({\it i.e.}, velocity) space of the photon\cite{15}.}.
This, therefore, implies that geometric phase differs from
dynamical phase and it involves global and topological information
on the time evolution of quantum systems. In addition to this
trigonometric geometric phase, there exists the so-called
hyperbolical geometric phase that is expressed by $2\pi(1-\cosh
\theta)$ with the hyperbolical cosine $\cosh
\theta=\frac{1}{2}[\exp(\theta)+\exp(-\theta)]$ in some
time-dependent quantum systems, {\it e.g.}, the two-level atomic
system with electric dipole-dipole interaction and the
harmonic-oscillator system\cite{22}. It is verified that the
generators of the Hamiltonians of these quantum systems form the
$SU(1,1)$ Lie algebra. Further analysis indicates that quantum
systems, which possess the non-compact Lie algebraic structure
(whose group parameters can be taken to be infinity) will present
the hyperbolical geometric phase, while quantum systems with
compact Lie algebraic structure will give rise to the
trigonometric geometric phase.

Since Lorentz group (describing the boosts of reference frames in
the space direction) in the special theory of relativity is also a
non-compact group, this leads us to consider the topological
properties associated with space-time. We take into account the
gravitational analogue to the magnetic charge\cite{6}, {\it i.e.},
gravitomagnetic charge that is the source of gravitomagnetic field
just as the case that {\it mass} ({\it gravitoelectric charge}) is
the source of gravitoelectric field ({\it i.e.}, Newtonian
gravitational field in the sense of weak-approximation). In this
sense, gravitomagnetic charge is also termed the {\it dual mass}.
It should be noted that the concept of the ordinary mass is of no
physical significance for the gravitomagnetic charge; it is of
interest to investigate the relativistic dynamics and
gravitational effects as well as geometric properties of this {\it
topological dual mass} (should such exist). From the point of view
of differential geometry, matter may be classified into two
categories: gravitoelectric matter and gravitomagnetic matter. The
former category possesses mass and constitutes the familiar
physical world, while the latter possesses dual mass that would
cause the non-analytical property of space-time metric. Einstein's
field equation of gravitation in general theory of relativity
governs the couplings of gravitoelectric matter (which possesses
mass) to gravity (space-time); accordingly, we should have a field
equation governing the interaction of dual matter with gravity. By
making use of the variational principle, the gravitational field
equation of gravitomagnetic matter can be obtained where the dual
Einstein's tensor is denoted by\cite{Shen1}
$\epsilon^{\alpha\lambda\sigma\tau}R^{\beta}_{\ast
\lambda\sigma\tau}-\epsilon^{\beta\lambda\sigma\tau}R^{\alpha}_{\ast
\lambda\sigma\tau}$ with $\epsilon^{\alpha\lambda\sigma\tau}$ and
$R^{\beta}_{\ast\lambda\sigma\tau}$ being the four-dimensional
Levi-Civita completely antisymmetric tensor and the Riemann
curvature tensor that describes the space-time curvature,
respectively.  By exactly solving this field equation, one can
show that the topological property of the solution
$g_{t\varphi}(r,\theta)$ can be illustrated as follows: solid
angle subtended by the curve shows the topological property of the
gravitomagnetic vector potential of a static gravitomagnetic
charge at the origin of the spherical coordinate system. Take the
gravitomagnetic vector potentials ${\bf g}=\left(0, 0,
\frac{2\mu}{4\pi}\cdot\frac{(1-\cos \theta)}{r\sin \theta}\right)$
in spherical coordinate system, then the loop integral,
$\oint_{\rm curve}{\bf g}\cdot{\rm d}{\bf l}=\mu(1-\cos \theta)$,
is proportional to the solid angle subtended by the curve with
respect to the origin. The same situations arise in the adiabatic
quantum geometric phase (Berry's quantum phase), which reflects
the global or topological properties of time evolution (or
parameters evolution) of quantum systems.. Such property is in
analogy with that of the geometric quantum phase in the
time-dependent spin-gravity coupling ({\it i.e.}, the interaction
between a spinning particle with gravitomagnetic field\cite{20})
and other quantum adiabatic processes\cite{5,10}. The topological
properties of gravitomagnetic charge (dual mass) may be shown in
terms of the global features of geometric quantum
phase\footnote{Analogy between {\it topological dual mass} and
{\it geometric quantum phase} are as follows: the dynamical
correspondences to both of them are respectively the {\it mass}
and the {\it dynamical phase}; gravitomagnetic moment results from
the mass current, which is also the dynamical physical quantity;
both reveal the global properties of physical systems. Comparison
between gravitomagnetic charge and geometric phase enables to show
the topological properties of the former. The reason why the
topological property is important lies in that the global
description of the physical phenomena is essential to understand
the world. It is of interest that dual matter may constitute a
dual world where dual mass abides by their own dynamical and
gravitational laws, which is somewhat different from the laws in
our world; for example, dual mass is acted upon by a
gravitomagnetic Lorentz force in Newton's gravitational
(gravitoelectric) field, and the static dual mass produces the
gravitomagnetic field rather than the Newton's gravitoelectric
field. }. It follows that the expression of gravitomagnetic
potential, $g_{t\varphi}(r,\theta)$, due to dual mass is exactly
analogous to that of the trigonometric geometric phase. In the
similar fashion, it is readily verified that the gravitomagnetic
potential, $g_{\theta\varphi}(r,t)$, is similar to that of the
hyperbolical geometric phase. This feature originates from the
fact that the Lorentz group is a non-compact group.

Although there is no evidence for the existence of this
topological dual mass at present, it is still essential to
consider this topological or global phenomenon in General
Relativity. It is believed that there would exist formation (or
creation) mechanism of gravitomagnetic charge in the gravitational
interaction, just as some prevalent theories provide the
theoretical mechanism of existence of magnetic monopole in various
gauge interactions\cite{23,24}. Magnetic monopole in
electrodynamics and gauge field theory has been discussed and
sought after for decades, and the existence of the 't
Hooft-Polyakov monopole solution\cite{23} has spurred new interest
of both theorists and experimentalists\cite{24,25,26}. As the
topological gravitomagnetic charge in the curved space-time, dual
mass is believed to give rise to such interesting situation
similar to that of magnetic monopole. If it is indeed present in
universe, dual mass will also lead to significant consequences in
astrophysics and cosmology. We emphasize that although the
gravitomagnetic vector potential produced by the gravitomagnetic
charge is the classical solution to the field equation, this kind
of topological gravitomagnetic monopoles may arise not as
fundamental entities in gravity theory, {\it e.g.}, it will behave
like a topological soliton.

Gravitomagnetic charge has some interesting relativistic quantum
gravitational effects\cite{27,20,Shen1}, {\it e.g.}, the
gravitational {\it anti-}Meissner effect, which  may serve as an
interpretation of the smallness of the observed cosmological
constant. In accordance with quantum field theory, vacuum
possesses infinite zero-point energy density due to the vacuum
quantum fluctuations; whereas according to Einstein's theory of
general relativity, infinite vacuum energy density yields the
divergent curvature of space-time, namely, the space-time of
vacuum is extremely curved. Apparently it is in contradiction with
the practical fact, since it follows from experimental
observations that the space-time of vacuum is asymptotically flat.
In the context of quantum field theory a cosmological constant
corresponds to the energy density associated with the vacuum and
then the divergent cosmological constant may result from the
infinite energy density of vacuum quantum fluctuations. However, a
diverse set of observations suggests that the universe possesses a
nonzero but very small cosmological constant\cite{28,29,30,31}.
How can we give a natural interpretation for the above paradox?
Here, provided that vacuum matter is perfect fluid, which leads to
the formal similarities between the weak-gravity equation in
perfect fluid and the London's electrodynamics of
superconductivity, we suggest a potential explanation by using the
cancelling mechanism via gravitational {\it anti-}Meissner
effect\footnote{Note that in London's electrodynamics for
superconductivity, the equation governing the magnetic fields in
superconductors is of the type: $\nabla^{2}{\bf B}=\lambda^{2}{\bf
B}$ with $\lambda$ being a real number. In cosmological perfect
fluid, however, the equation governing the gravitomagnetic fields
in fluid is of the type: $\nabla^{2}{\bf B_{\rm g}}=-\lambda_{\rm
g}^{2}{\bf B_{\rm g}}$, which means the absence of the
gravitational analogue to the superconducting Meissner's effect.
We can refer to this phenomenon as the gravitational
(gravitomagnetic) {\it anti-}Meissner effect due to the minus sign
on the right-handed side of the latter equation.}: the
gravitoelectric field (Newtonian field of gravity) produced by the
gravitoelectric charge (mass) of the vacuum quantum fluctuations
is  exactly cancelled by the gravitoelectric field due to the
induced current of the gravitomagnetic charge of the vacuum
quantum fluctuations; the gravitomagnetic field produced by the
gravitomagnetic charge (dual mass) of the vacuum quantum
fluctuations is exactly cancelled by the gravitomagnetic field due
to the induced current of the gravitoelectric charge (mass
current) of the vacuum quantum fluctuations. Thus, at least in the
framework of weak-field approximation, the extreme space-time
curvature of vacuum caused by the large amount of the vacuum
energy does not arise, and the gravitational effects of
cosmological constant is eliminated by the contributions of the
gravitomagnetic charge (dual mass). If gravitational Meissner{\it
anti-}Meissner effect is of really physical significance, then it
is necessary to apply this effect to the early universe where
quantum and inflationary cosmologies dominate the evolution of the
Universe.
\section{$\mathcal Concluding$ $\mathcal remarks$}
For the present, it is possible to investigate quantum mechanics
in weak-gravitational fields, with the development of detecting
and measuring technology such as laser-interferometer technology
and so on. These investigations enable physicists to test validity
or universality of fundamental laws and principles of General
Relativity in, for instance, the microscopic quantum regimes. The
above-mentioned effects associated gravitomagnetic fields reflect
the relativistic quantum gravitational properties of
matter-gravity couplings in weak-gravitational fields. It is
believed that these effects will give rise to further interest of
investigation, since there exist many physically potential
applications of these relativistic quantum gravitational effects
in quantum optics, gravity theory, quantum theory, cosmology,
applied physics and other related areas as well.

Study of the geometric property in quantum regimes is an
interesting and valuable direction. Since it reveals the global
and topological properties of evolution of quantum systems,
geometric phase has many applications in various branches of
physics, say, in the coupling of neutron spin to the Earth's
rotation\cite{27,32}, a potential application may be suggested
where the information on the Earth's variations of rotating
frequency will be obtained by measuring the geometric phase of the
oppositely polarized neutrons through the {\it neutron-gravity
interferometer experiment}. The topological charge in curved
space-time also deserves further investigation, since it reflects
plentiful global or geometric properties hidden in the gravity
theory. It is believed that both theoretical and experimental
interest in this direction may enables people to understand the
global phenomena of the physical world better.

\end{document}